\documentclass[epj]{webofc}
\usepackage[varg]{txfonts}   
\wocname{EPJ Web of Conferences}
\woctitle{INPC 2013}
\begin{document}
\title{Profiling hot and dense nuclear medium with high transverse momentum
hadrons produced in d+Au and Au+Au collisions by the PHENIX experiment
at RHIC}
%
%

\author{Takao Sakaguchi\inst{1}\fnsep\thanks{\email{takao@bnl.gov}}, for the PHENIX collaboration} 

\institute{Brookhaven National Laboratory, Physics Department, Upton, NY 11973--5000, USA}

\abstract{%
PHENIX measurements of high transverse momentum ($p_T$) identified hadrons
in $d$+Au and Au+Au collisions are presented. The nuclear modification
factors ($R_{d{\rm A}}$ and $R_{\rm AA}$) for $\pi^0$ and $\eta$ are
found to be very consistent in both collision systems, respectively.
Using large amount of $p+p$ and Au+Au datasets, the fractional momentum
loss ($S_{\rm loss}$) and the path-length dependent yield of $\pi^0$ in
Au+Au collisions are obtained.
The hadron spectra in the most central $d$+Au and the most peripheral
Au+Au collisions are studied. The spectra shapes are found to be similar
in both systems, but the yield is suppressed in the most peripheral
Au+Au collisions.
}
\maketitle
\section{Introduction}
\label{intro}
The interaction of hard scattered partons with the medium created
by heavy ion collisions (i.e., quark-gluon plasma, QGP) has been of
interest since the beginning of the RHIC running~\cite{Wang:1998bha}.
A large suppression of the yields of high transverse momentum ($p_T$)
hadrons which are the fragments of such partons was observed,
suggesting that the matter is sufficiently dense to cause
parton-energy loss prior to hadronization~\cite{Adler:2003qi}.
Absence of the hadron suppression in $d$+Au collisions supported the
parton-energy loss scenario~\cite{Adler:2003ii}. After accumulating
a large amount of $p+p$, $d$+Au, and Au+Au collision events,
we substantially extended the degree of freedom in high $p_T$ hadron
measurements. In this paper, we show the recent studies of the QGP
using high $p_T$ hadrons by the PHENIX experiment.

\section{$\pi^0$ and $\eta$ measurements in d+Au and Au+Au collisions}
\label{sec-2}
\begin{figure}[h]
\centering
\includegraphics[width=11cm]{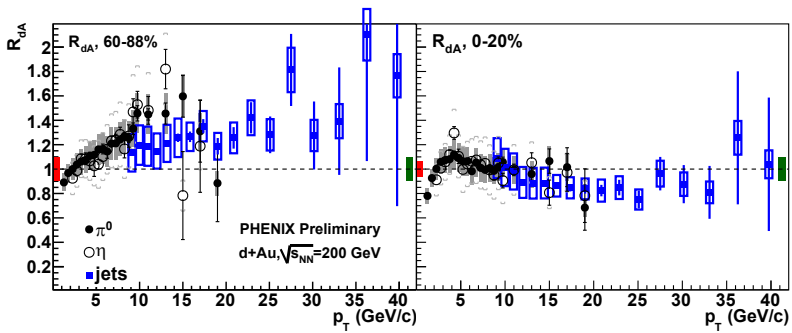}
\label{fig1}       
\caption{$R_{d{\rm A}}$ for $\pi^0$, $\eta$ and fully-reconstructed jets in $d$+Au collisions.}
\end{figure}

The PHENIX experiment~\cite{Adcox:2003zm} has been exploring the highest
$p_T$ region with single $\pi^0$ and $\eta$ mesons. They are leading
hadrons of jets, and thus provide a good measure of momentum of hard
scattered partons. Here, we present the results obtained from
$d$+Au collisions collected in the RHIC Year-2008 run
(80 nb$^{-1}$) and Au+Au collisions in the Year-2007 run
(0.81 nb$^{-1}$).
Figure~\ref{fig1} shows the nuclear modification factors
 ($R_{d{\rm A}}\equiv(dN_{d{\rm A}}/dydp_{T})/(\langle T_{d{\rm A}}\rangle d\sigma_{pp}/dydp_{T})$)
 for $\pi^0$, $\eta$ and fully-reconstructed jets in $d$+Au
collisions at $\sqrt{s_{NN}}$=200\,GeV.
They are very consistent each other, and also consistent with unity
at low $p_T$ in both most central and peripheral collisions.
However, at high $p_T$, the yields are suppressed
in most central collisions and enhanced in most peripheral collisions.

\begin{figure}[h]
\begin{minipage}{60mm}
\centering
\includegraphics[width=6.0cm]{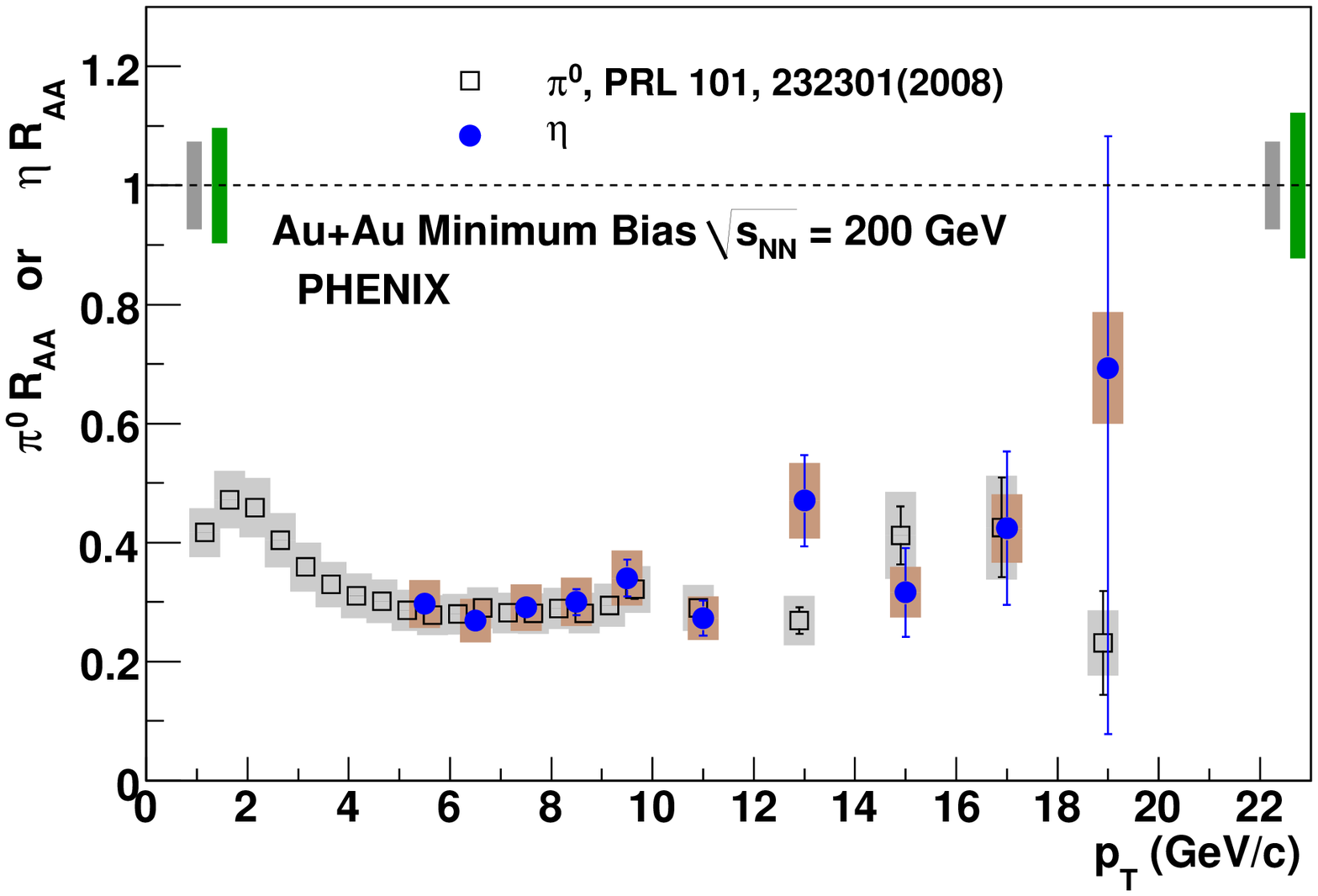}
\end{minipage}
\begin{minipage}{80mm}
\centering
\includegraphics[width=8.0cm]{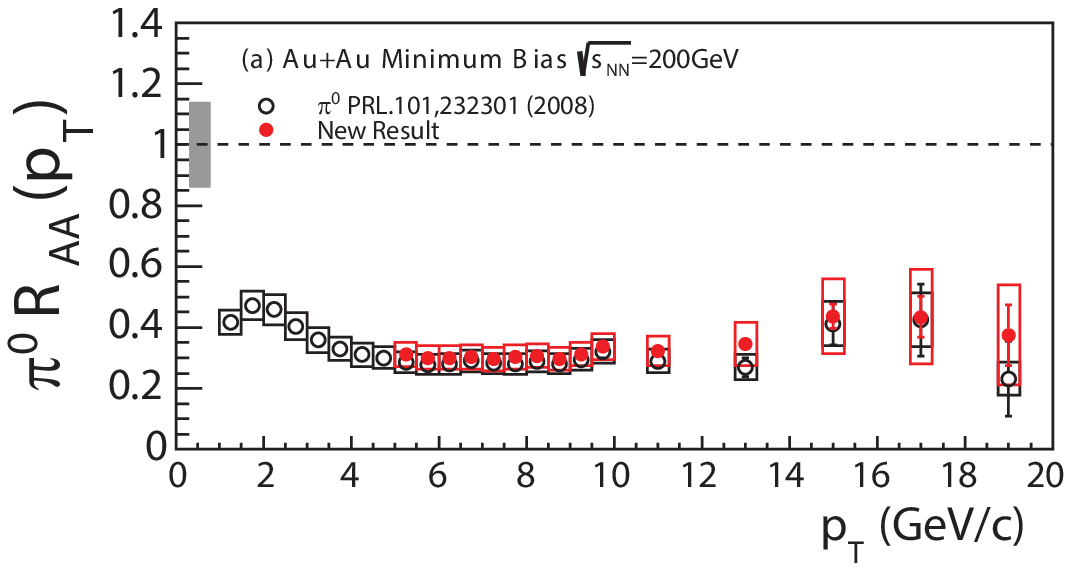}
\end{minipage}
\label{fig2}       
\caption{ (a, left) $R_{\rm AA}$ for $\pi^0$ and $\eta$ in minimum bias Au+Au
collisions. (b, right) $R_{\rm AA}$ for $\pi^0$ from the RHIC Year-2004 run
and Year-2007 run.}
\end{figure}

The consistency of $\pi^0$ and $\eta$ are also seen in $R_{\rm AA}$ ($\equiv(dN_{\rm AA}/dydp_{T})/(\langle T_{\rm AA}\rangle d\sigma_{pp}/dydp_{T})$) in 200\,GeV Au+Au collisions as shown
in Figure~\ref{fig2}(a)~\cite{Adare:2010dc}.
Because $\eta$ has four times larger mass compared to that of $\pi^0$,
one can resolve two photons decaying from $\eta$ up to four times
larger $p_T$ of $\pi^0$, resulting in a higher $p_T$ reach with smaller
systematic errors with $\eta$.
Figure~\ref{fig2}(b) demonstrates that the $\pi^0$ from the Year-2007
run has smaller errors and is consistent with that from the Year-2004
run~\cite{Adare:2012wg}.

The recent result of single electron measurement shows that the $R_{d{\rm A}}$
and $R_{\rm AA}$ for light hadrons and electrons from heavy flavor
hadrons have similar trend of enhancement and suppression, except
for low $p_T$ region, where soft production is dominant~\cite{Adare:2012qb}.
This fact suggests that the interaction of light hadrons and
heavy hadrons with medium has same system dependence.

\section{Fractional momentum loss of hadrons in Au+Au collisions}
\label{sec-3}
\begin{figure}[h]
\begin{minipage}{45mm}
\centering
\includegraphics[width=4.5cm]{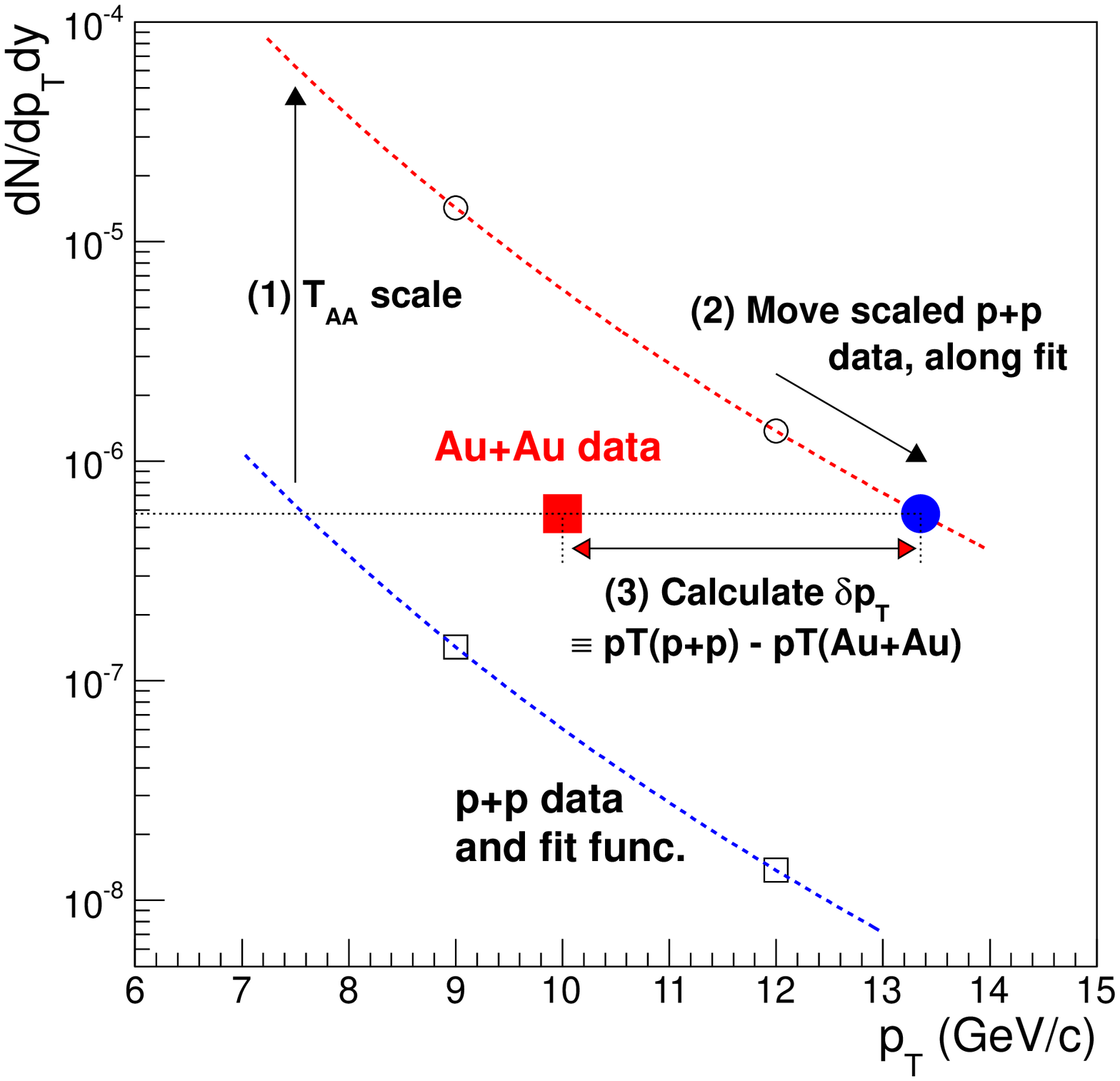}
\end{minipage}
\begin{minipage}{40mm}
\centering
\vspace{5mm}
\includegraphics[width=4.0cm]{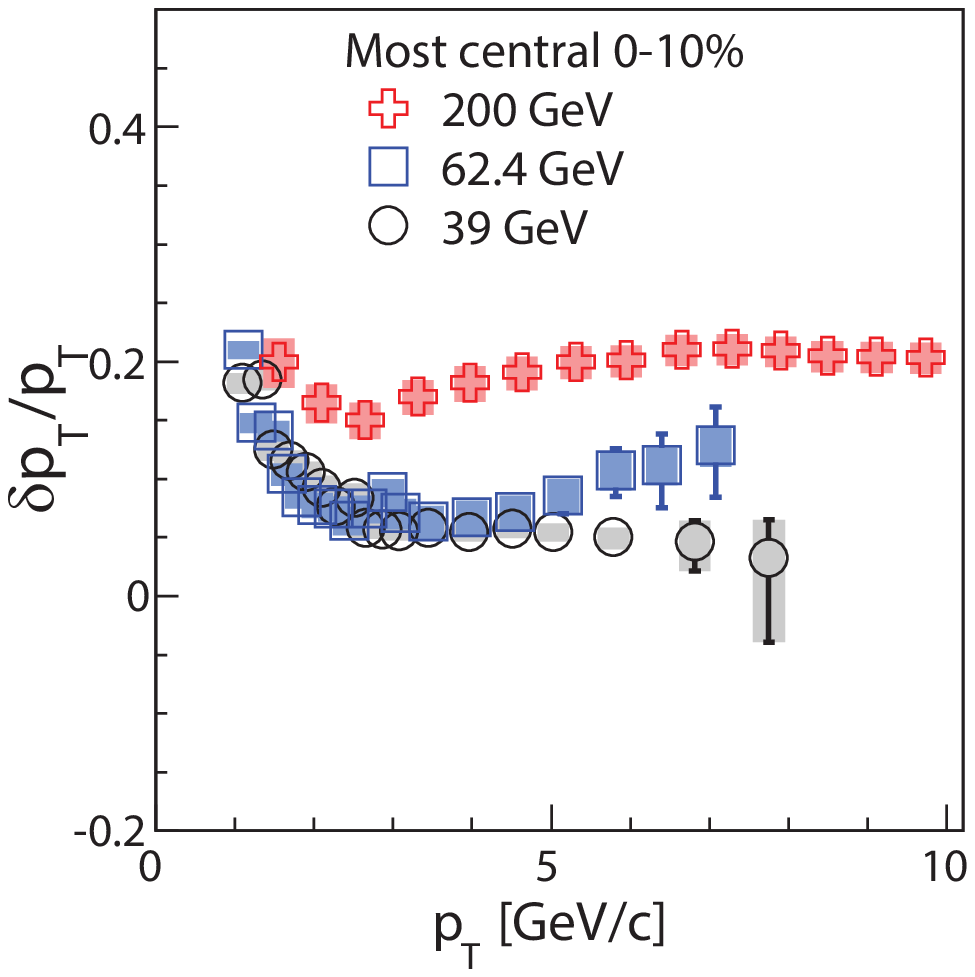}
\end{minipage}
\begin{minipage}{55mm}
\centering
\includegraphics[width=5.5cm]{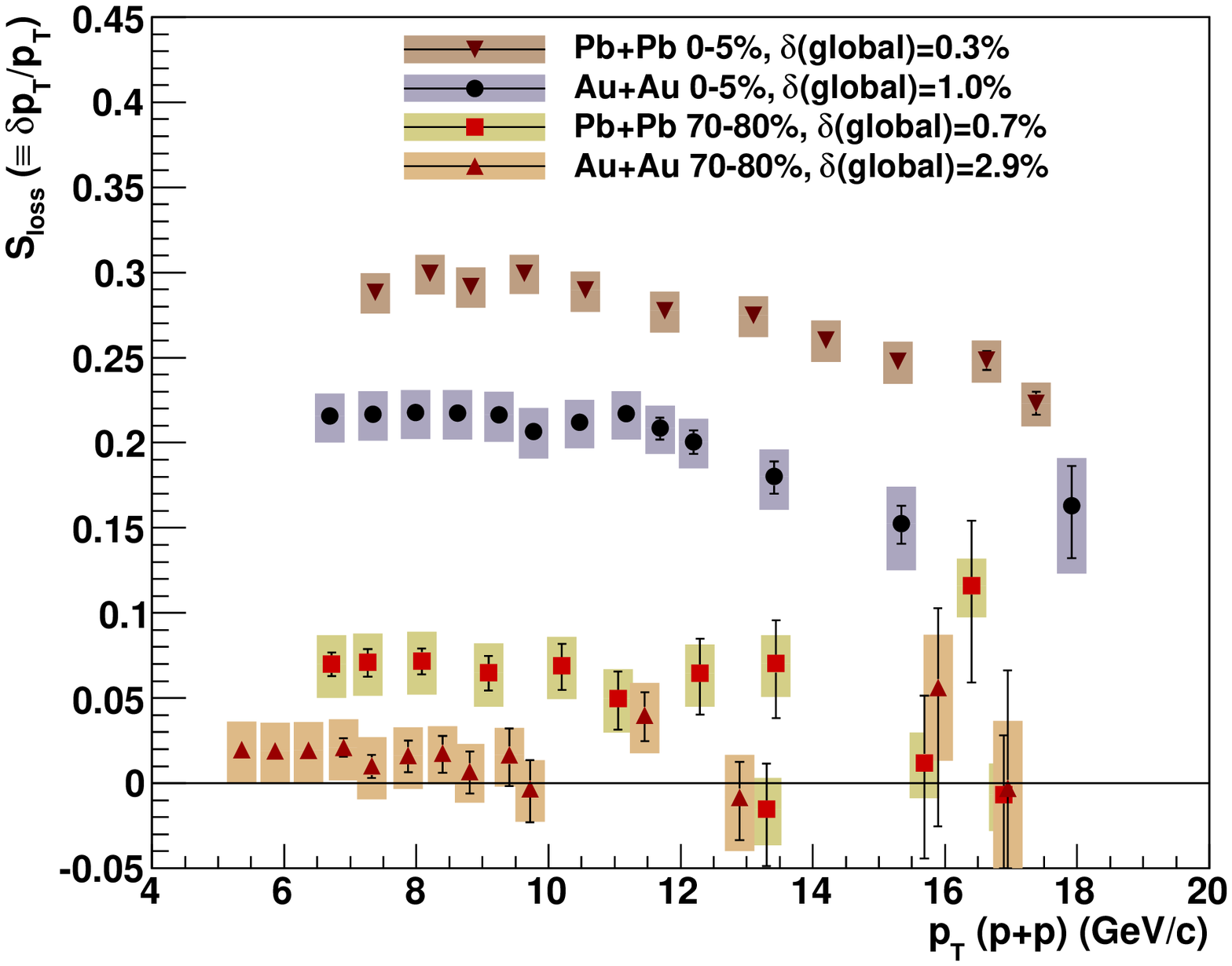}
\end{minipage}
\label{fig3}       
\caption{(a, left) Method of calculating average $S_{\rm loss}$. (b, middle)
$S_{\rm loss}$ for $\pi^0$ for 0-10\,\% centrality 39, 62, and 200\,GeV
Au+Au collisions. (c, right) $S_{\rm loss}$ for $\pi^0$ in 200\,GeV Au+Au
collisions and charged hadrons in 2.76\,TeV Pb+Pb collisions.}
\end{figure}

The large amount of events collected in $p+p$ and Au+Au collisions
made us possible to quantify the energy loss effect from a different
aspect. Experiments have been looking at the suppression of the yield
to see the effect. However, the suppression is primarily the consequence
of the reduction of momentum of hadrons which have exponential $p_T$
distributions.  We have statistically extracted the fractional
momentum loss ($S_{\rm loss}\equiv \delta p_T/p_T$) of the partons
using the hadron $p_T$ spectra measured in $p+p$ and Au+Au
collisions~\cite{Adare:2012wg}.
Figure~\ref{fig3}(a) depicts the method to compute the $S_{\rm loss}$.
Using this method, we computed the $S_{\rm loss}$ in Au+Au collisions at
$\sqrt{s_{NN}}=$39, 62, and 200\,GeV as shown in
Figure~\ref{fig3}(b)~\cite{Adare:2012uk}.
We also computed the $S_{\rm loss}$ in 2.76\,TeV Pb+Pb collisions using charged
hadron spectra measured by the ALICE experiment~\cite{Aamodt:2010jd}
as shown in Figure~\ref{fig3}(c). $S_{\rm loss}$'s vary by a factor of six
from 39\,GeV Au+Au to 2.76\,TeV Pb+Pb collisions.

\section{Path-length and collision system dependence of parton energy loss}
\begin{figure}[h]
\centering
\includegraphics[width=12cm]{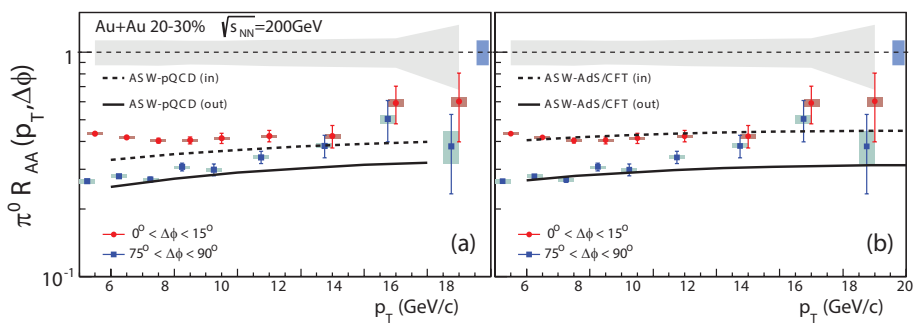}
\caption{$R_{\rm AA}(\phi)$ of $\pi^0$ in 20–30\,\% centrality for in-plane
 and out-of-plane. (a, left) Data are compared with a pQCD-inspired model,
and (b, right) an AdS/CFT-inspired model.}
\label{fig4}       
\end{figure}

With larger statistics, we were able to measure the $R_{\rm AA}$ of $\pi^0$ for
in- and out-of event planes. Figure~\ref{fig4} shows the ones for $\pi^0$s
in 20-30\,\% central 200\,GeV Au+Au
collisions~\cite{Adare:2012wg}.
The difference of the yield provides path-length dependence of
yield modification. Depending on the energy loss models, the
powers of the path-length dependence change.
The data favors an AdS/CFT-inspired (strongly coupled) model
rather than pQCD-inspired (weakly coupled) model, implying that
the energy loss is $L^3$ dependent rather than $L^2$ dependence,
where $L$ denotes the path-length of partons in the medium.

\begin{figure}[h]
\centering
\sidecaption
\includegraphics[width=7.0cm]{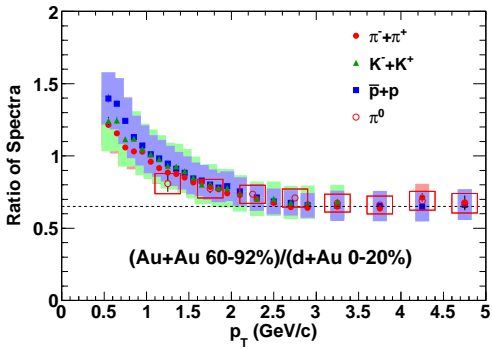}
\caption{Ratio of invariant yield of particles in peripheral Au+Au
 (60–92\,\%) to central $d$+Au (0–20\,\%) collisions as a function of $p_T$.}
\label{fig5}       
\end{figure}

We note that the $N_{\rm coll}$ and $N_{\rm part}$ values are quite
consistent in certain central $d$+Au and peripheral Au+Au collisions.
The ratio of $N_{\rm coll}$ in 0-20\% $d$$+$Au to that
in 60-92\% Au$+$Au is 1.02~$\pm$~0.22, and the same ratio for
$N_{\rm part}$ values is 1.04~$\pm$~0.21.
Motivated by this fact, we took the ratio of the spectra in 60-92\,\%
Au$+$Au to 0-20\,\% $d$$+$Au collisions for identified particles
as shown in Figure~\ref{fig5}~\cite{Adare:2013esx}.
The ratios tend to the same value of roughly 0.65 for each particle
species at and above 2.5-3\,GeV/$c$.
This universal scaling is strongly suggestive of a
common particle production mechanism between peripheral Au$+$Au and central
$d$$+$Au collisions. The trend of overall rise in low $p_T$ may come from
rapidity shift in asymmetric collisions in $d$+Au. There is also mass
dependence of the rise seen in lower $p_T$.
Assuming that the cold nuclear effect scales with $N_{\rm coll}$
or $N_{\rm part}$, the ratio 0.65 may be attributable to the parton
energy loss in peripheral Au$+$Au collisions.

\section{Summary}
PHENIX measurement of high $p_T$ identified hadrons
in $d$+Au and Au+Au collisions are presented. The $R_{\rm AA}$ for
$\pi^0$ and $\eta$ are found to be very consistent in both
collision systems, respectively. The $S_{\rm loss}$'s of high $p_T$ hadrons are
computed from 39\,GeV Au+Au over to 2.76\,TeV Pb+Pb, and found that they
vary by a factor of six. The path-length dependent $\pi^0$ yield
deduced that the energy loss of partons is $L^3$ dependent. 
It was found that the hadron production mechanism in central $d$+Au
and peripheral Au+Au is similar, but the ratio of the yields is
$\sim$0.65 which may be attributable to the parton
energy loss in peripheral Au$+$Au collisions.


%

\begin{thebibliography}{99}
%
%
\bibitem{Wang:1998bha} 
  X.~-N.~Wang,
  Phys.\ Rev.\ C {\bf 58}, 2321 (1998).
\bibitem{Adler:2003qi} 
  S.~S.~Adler {\it et al.}  [PHENIX Collaboration],
  Phys.\ Rev.\ Lett.\  {\bf 91}, 072301 (2003).
\bibitem{Adler:2003ii} 
  S.~S.~Adler {\it et al.}  [PHENIX Collaboration],
  Phys.\ Rev.\ Lett.\  {\bf 91}, 072303 (2003).
\bibitem{Adcox:2003zm} 
  K.~Adcox {\it et al.}  [PHENIX Collaboration],
  Nucl.\ Instrum.\ Meth.\ A {\bf 499}, 469 (2003).
\bibitem{Adare:2010dc} 
  A.~Adare {\it et al.}  [PHENIX Collaboration],
  Phys.\ Rev.\ C {\bf 82}, 011902 (2010).
\bibitem{Adare:2012wg} 
  A.~Adare {\it et al.}  [PHENIX Collaboration],
  Phys.\ Rev.\ C {\bf 87}, 034911 (2013).
\bibitem{Adare:2012qb} 
  A.~Adare {\it et al.}  [PHENIX Collaboration],
  Phys.\ Rev.\ Lett.\  {\bf 109}, 242301 (2012).
\bibitem{Adare:2012uk} 
  A.~Adare {\it et al.}  [PHENIX Collaboration],
  Phys.\ Rev.\ Lett.\  {\bf 109}, 152301 (2012).
\bibitem{Aamodt:2010jd} 
  K.~Aamodt {\it et al.}  [ALICE Collaboration],
  Phys.\ Lett.\ B {\bf 696}, 30 (2011).
\bibitem{Adare:2013esx} 
  A.~Adare {\it et al.}  [PHENIX Collaboration],
  arXiv:1304.3410 [nucl-ex], in press.
\end{thebibliography}
%
%

\end{document}